\documentclass[aps,prd,twocolumn]{revtex4}
\usepackage[pdftex]{graphicx}
\usepackage{amsmath,amssymb,bm}
\usepackage[linkcolor=blue,citecolor=blue]{hyperref} 
\hypersetup{colorlinks=true}
\newcommand\ee{\end{equation}}
\newcommand\be{\begin{equation}}

\begin{document}

\title{Atmospheric Neutrinos in a Next-Generation Xenon Dark Matter Experiment}
\author{Jayden~L.~Newstead$^{\bf a,b,d}$}
\author{Rafael~F.~Lang$^{\bf b}$}
\author{Louis~E.~Strigari$^{\bf c}$}
\affiliation{$^{\bf a}$Department of Physics, Arizona State University, Tempe, AZ 85287, USA}
\affiliation{$^{\bf b}$Department of Physics and Astronomy, Purdue University, West Lafayette, IN 47907, USA}
\affiliation{$^{\bf c}$Mitchell Institute for Fundamental Physics and Astronomy,  Department of Physics and Astronomy, Texas A\&M University, College Station, TX 77845, USA}
\affiliation{$^{\bf d}$School of Physics, The University of Melbourne, Victoria 3010, Australia}

\begin{abstract}
We study the sensitivity of future xenon- and argon-based dark matter and neutrino detection experiments to low-energy atmospheric neutrinos. Not accounting for experimental backgrounds, the primary obstacle for identifying nuclear recoils induced by atmospheric neutrinos is the tail of the electron recoil distribution due to $pp$ solar neutrinos. We use the NEST code to model the solar and atmospheric neutrino signals in a xenon detector and find that an exposure of 700 tonne-years will produce a $5\sigma$ detection of atmospheric neutrinos. We explore the effect of different detector properties and find that a sufficiently long electron lifetime is essential to the success of such a measurement.
\end{abstract}

\maketitle

\section{Introduction} 

Multi-ton liquid noble dark matter direct detection experiments will soon be sensitive to coherent neutrino-nucleus elastic scattering (CE$\nu$NS) from astrophysical neutrinos, specifically from the Sun, the atmosphere~\cite{Billard:2013qya,Billard:2014yka}, and possibly from Galactic~\cite{Lang:2016zhv} and diffuse supernovae~\cite{Strigari:2009bq}. Identifying these neutrinos is an important goal for neutrino physics~\cite{Dutta:2019oaj}, and is an important milestone for future multi-purpose dark matter detectors~\cite{Aalbers:2016jon,Schumann:2015cpa}. 

For spin-independent dark matter-nucleus interactions, the nuclear recoil spectrum from a $\sim$~6~GeV dark matter particle with cross section $\lesssim 10^{-45}$~cm$^2$ mimics the spectrum from the $^8$B component of the solar neutrino flux. The detailed sensitivity to solar neutrinos has been the subject of several studies, from the perspective of both a neutrino signal and a background to dark matter detection~\cite{Billard:2013qya,Billard:2014yka,Dent:2016wor,Dent:2016iht,OHare:2020lva,Newstead:2018muu}. For atmospheric neutrinos, a $\sim$~100~GeV dark matter particle with cross section $\lesssim 10^{-48}$~cm$^2$ will mimic the recoil spectrum from atmospheric neutrinos. However, because of the larger exposures required to gain sensitivity to atmospheric neutrinos, understanding this component as either a neutrino signal or a dark matter background has been subject to less scrutiny in the literature. 

In this paper, we undertake the first experimental-based study of multi-ton scale dark matter and neutrino experiments based on liquid xenon or argon to atmospheric neutrinos. We focus on identifying the neutrino-induced nuclear recoil signal in the presence of backgrounds, with the main one arising from electron recoils from the $pp$ component of the solar neutrino flux. Previous studies have estimated the atmospheric neutrino event rate to be order unity for $\sim 20$ tonne-year scale exposures~\cite{Strigari:2009bq,Billard:2013qya}. Here, we extend upon these results and perform simulations of the nuclear recoil signal induced by atmospheric neutrinos in a realistic detection configuration using the Noble Element Simulation Technique (NEST) package~\cite{Szydagis:2011tk,szydagis_m_2019_3357973}.

This paper is organized as follows. In Section II we discuss the characteristics of the atmospheric neutrino signal in future xenon experiments. In Section III we discuss the NEST simulation and present our statistical method for determining the significance of the neutrino signal as a function of exposure. In Section IV we present our primary results, and conclude in Section V. 

\section{Atmospheric neutrino signal}

The atmospheric neutrino flux has been calculated down to energies of 10~MeV using the FLUKA code~\cite{Battistoni:2002ew}. At these energies, the flux originates mostly from pion decay, so that the flavor composition is $\sim2/3$ muon flavor and $\sim 1/3$ electron flavor. The sub-GeV normalization of the atmospheric neutrino flux has not been directly measured, and there are theoretical uncertainties that arise from several physical processes. One such uncertainty arises from the fact that the cosmic ray flux at the top of the Earth's atmosphere differs from the cosmic ray flux in the interstellar medium. A second uncertainty is from the geomagnetic field, which induces a cut-off in the low-energy cosmic ray spectrum. Detailed modeling of both of these effects implies that for energies $\lesssim 100$~MeV, the uncertainty on the predicted atmospheric neutrino flux is approximately 20\%~\cite{Honda:2011nf}. Due in particular to the cutoff in the rigidity of cosmic rays induced by the Earth's geomagnetic field at low energies, the atmospheric neutrino flux is larger for detectors that are nearer to the poles~\cite{Honda:2011nf}. 

\begin{figure}{ht}
    \centering
    \includegraphics[width=0.9\columnwidth]{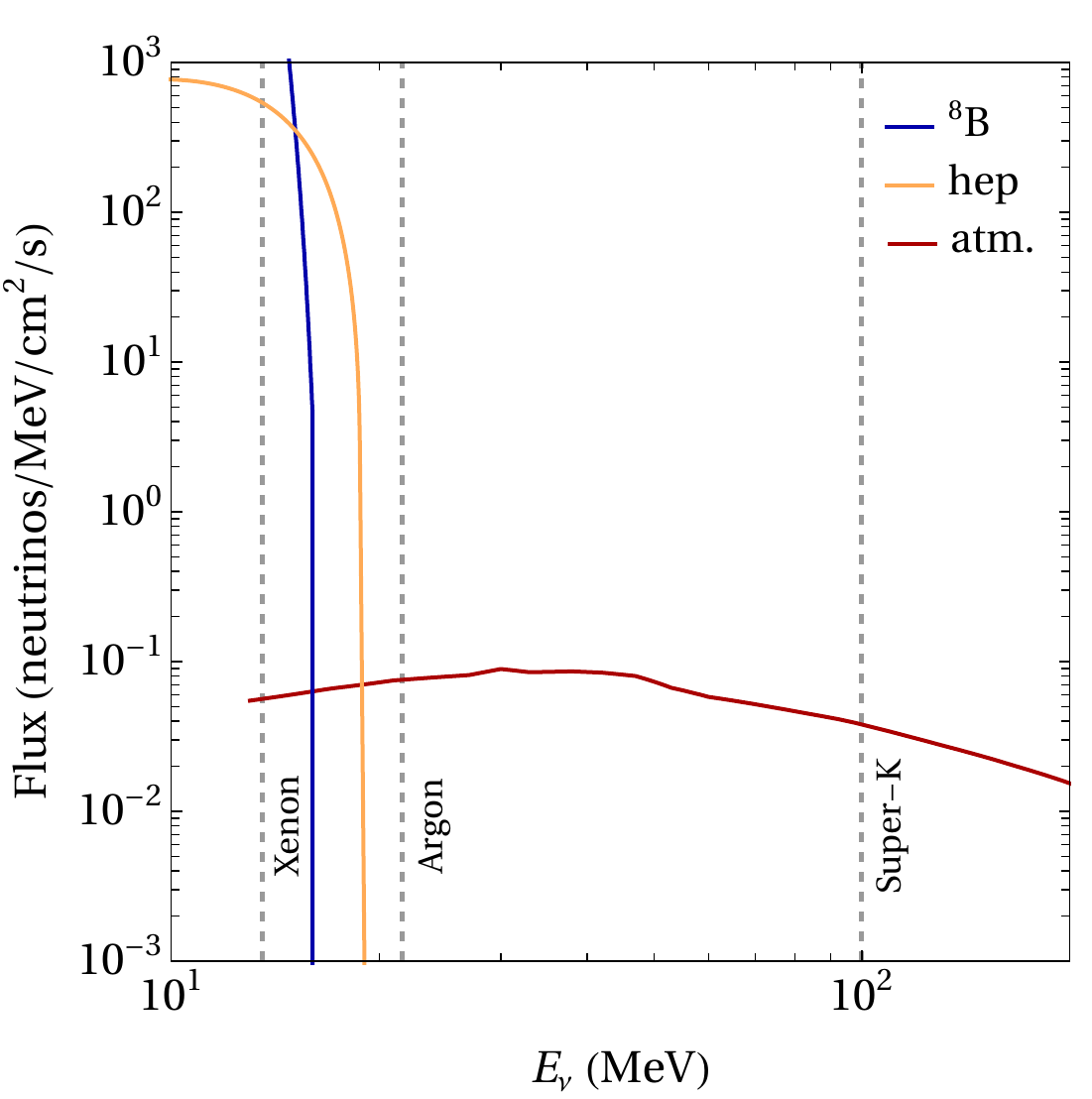}
    \caption{The solar and atmospheric neutrino flux in the $E_\nu$=10--100~MeV range. The vertical dashed lines show the corresponding neutrino energies above which xenon and argon detectors are sensitive to, given their nuclear recoil energy thresholds ($E_R=3$~keV for xenon, $E_R=25$~keV for argon). Also shown is the neutrino energy range above which Super-K is sensitive to~\cite{Richard:2015aua}.}
    \label{fig:flux}
\end{figure}

In Fig.~\ref{fig:flux}, we show the total atmospheric neutrino flux, summing over flavor and matter/anti-matter, at average solar activity, for the latitude of the Gran Sasso underground laboratory (LNGS). Since LNGS is a possible location for future Generation-3 xenon or argon experiments, and since such experiments are expected to run for more than a decade, we use this flux as shown in Fig.~\ref{fig:flux} for all following calculations. 

Atmospheric neutrinos interact via CE$\nu$NS in a xenon or argon dark matter detector. CE$\nu$NS is a standard model neutral-current process whose differential cross section can be calculated as:
\be
\frac{d\sigma}{d E_R} = \frac{G_F^2 m_T }{\pi}Q_w^2\left(1-\frac{m_T E_R}{2E_\nu^2}\right)F^2(E_R),
\ee
where $G_F$ is Fermi constant, $m_T$ and $Q_w$ are the mass and weak charge of the target nuclei, $E_R$ is the nuclear recoil energy and $E_\nu$ is the incoming neutrino energy. The form factor, $F(E_R)$, accounts for the loss of coherence at larger momentum transfers and is the dominant source of uncertainty in this cross section, contributing around 5\% to the rate normalization for neutrinos in this energy range. Here, we take the form factor to be the one proposed by Helm~\cite{Helm:1956zz}. The coherent nature of the interaction implies a scaling of the cross section with the number of nucleons squared, but the relatively small weak charge of the proton means that the scaling is closer to the number of neutrons squared. This implies that large atomic mass targets are favored for their neutron-rich nuclei.

\begin{figure}
    \centering
    \includegraphics[width=0.9\columnwidth]{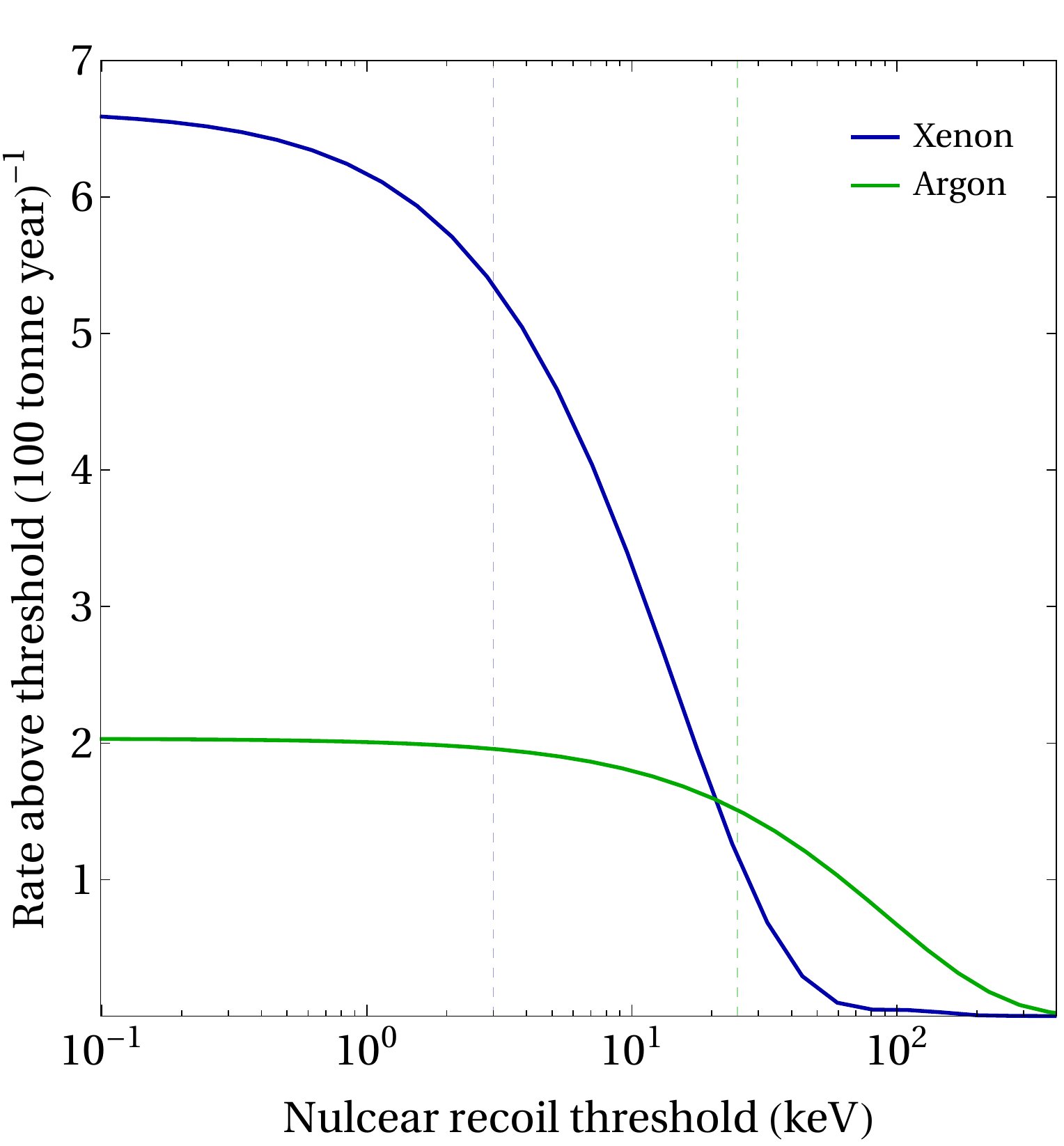}
    \caption{The integrated event rate of atmospheric neutrino CE$\nu$NS above a given threshold in xenon (blue) and argon (green) detector targets. The vertical dashed lines correspond to illustrative thresholds of 3~keV and 25~keV for xenon and argon, respectively.}
    \label{fig:rates}
\end{figure}

The differential event rate per unit detector mass can be calculated from
\be
\frac{d^2 R}{d E_\nu\, d E_R} = \frac{1}{m_T}\frac{d\sigma}{d E_R}\frac{d\phi_{\nu,i}}{d E_\nu}\, \Theta\left(E_{R,\mathrm{max}}(E_\nu)-E_R\right)
\label{eq:diffRate}
\ee
where $\phi_{\nu,i}$ is the $i$th neutrino flux, and $\Theta$ is the Heaviside step function which restricts $E_R$ to be less than the maximum value, corresponding to back-to-back scattering:
\be
E_{R,\mathrm{max}} =  \frac{2 E_\nu^2}{m_T+2E_\nu^2}.
\ee
The total event rate in an energy bin can then be obtained by integrating Eq.~\ref{eq:diffRate} over the relevant $E_R$ and $E_\nu$. The dependence on neutron number is exhibited by comparing the CE$\nu$NS rate for xenon and argon targets, as shown in Fig.~\ref{fig:rates}, where we have calculated the total rate above a given threshold. The CE$\nu$NS rate falls sharply with increasing recoil energy due to both loss of coherence and kinematic phase space, highlighting the necessity of a low detector threshold.

\begin{figure}
    \centering
    \includegraphics[width=0.9\columnwidth]{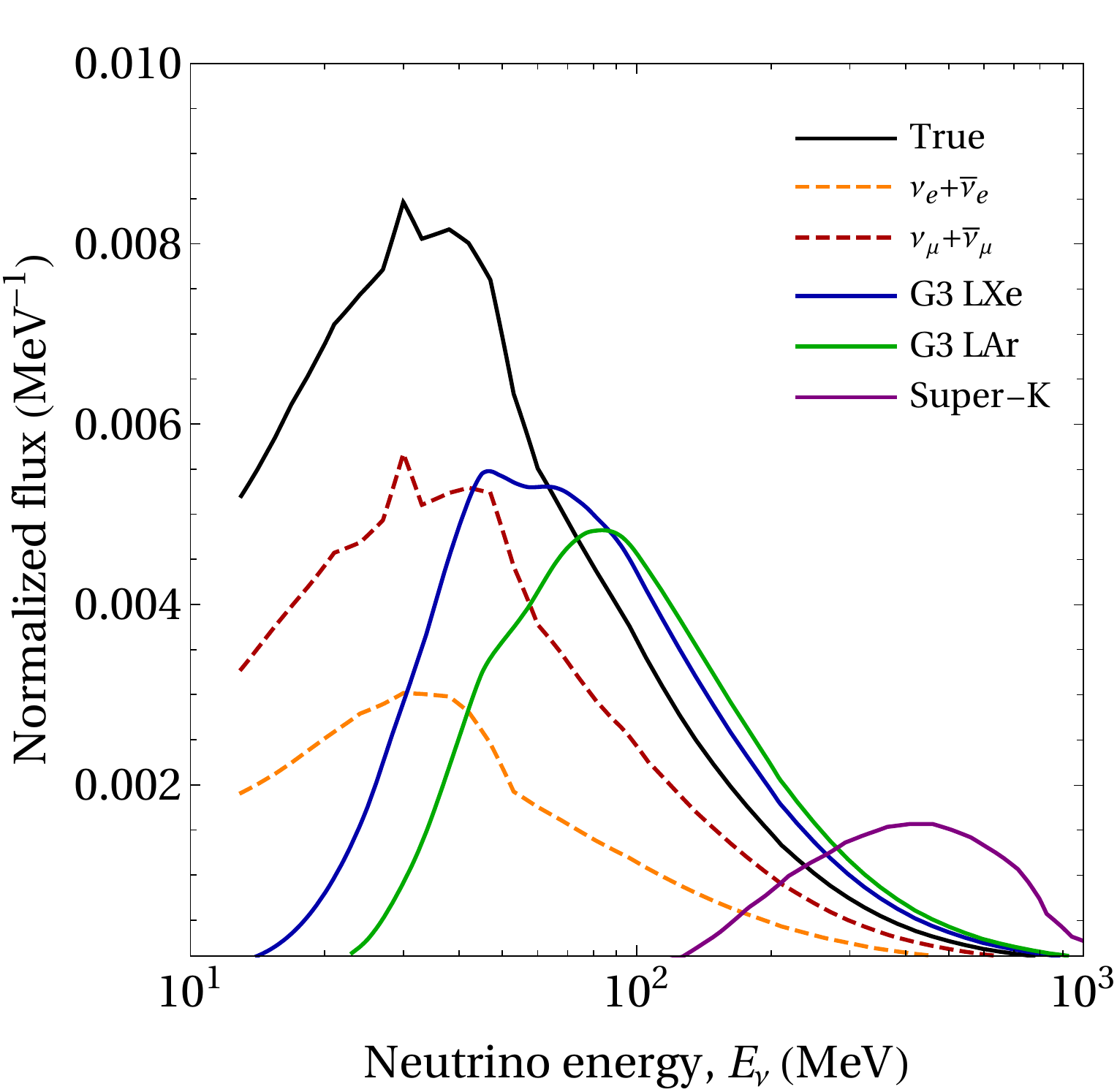}
    \caption{The differential fluxes of atmospheric neutrinos that are accessible by various experiments, normalized to unity. The electron and muon-flavored fluxes are indicated with the dashed curves, and the solid black curve is the total atmospheric neutrino flux, summed over all flavors. The features in the neutrino fluxes result from pion and muon decay at rest~\cite{Peres:2009xe}. Future dark matter experiments will access an atmospheric neutrino energy range that is not accessible to Super-K.}
    \label{fig:fluxComp}
\end{figure}

From the kinematic limits we can find that a detector sensitive to nuclear recoils in the energy range $\sim 1-50$~keV will be sensitive to neutrinos in the energy range $\sim 40-60$~MeV. More precisely, we can asses the range of energies of atmospheric neutrinos a given detector is sensitive to by integrating Eq.~\ref{eq:diffRate} over $E_R$ above a specified threshold. The result of this integration as a function of $E_\nu$ is given in Fig~\ref{fig:fluxComp}, indicating the neutrino energy range that a xenon and argon detector with $E_R \geqslant 3$~keV and $E_R \geqslant 25$~keV (respectively) would be sensitive to. For comparison we also show the lowest energy channel (sub-GeV single-ring electron-like events) that Super-Kamiokande was sensitive to in their atmospheric neutrino analysis~\cite{Richard:2015aua}. As indicated, Super-Kamiokande is sensitive to neutrinos $\gtrsim 100$~MeV for their fully-contained electron-like events. We note that JUNO would be sensitive to low-energy atmospheric neutrinos through the charged current channel~\cite{An:2015jdp}.

\section{Method}


\subsection{Detector properties}

Dual-phase noble time-projection chambers (TPCs) have proven to be a robust and scalable detector design for direct dark matter searches~\cite{Aprile:2018dbl,Cui:2017nnn,Akerib:2016vxi}. Detectors of this design are sensitive to $\mathcal{O}$(1~keV) nuclear recoils and provide 3D position reconstruction of events. The position reconstruction allows for detector fiducialization, where one takes advantage in particular of xenon detectors to self-shield, to achieve very low background in the central target volume. This is achieved through the detection of both the scintillation photons and ionization electrons that are produced by interactions in the detector bulk. The prompt scintillation light signal, S1, is measured directly by an array of PMTs on the top and bottom of the detector. The liberated electrons are drifted to the surface of the liquid phase and extracted into the gas phase where, through an avalanche process, they produce the delayed scintillation signal, S2.

\begin{figure*}
    \includegraphics[width=0.40\linewidth]{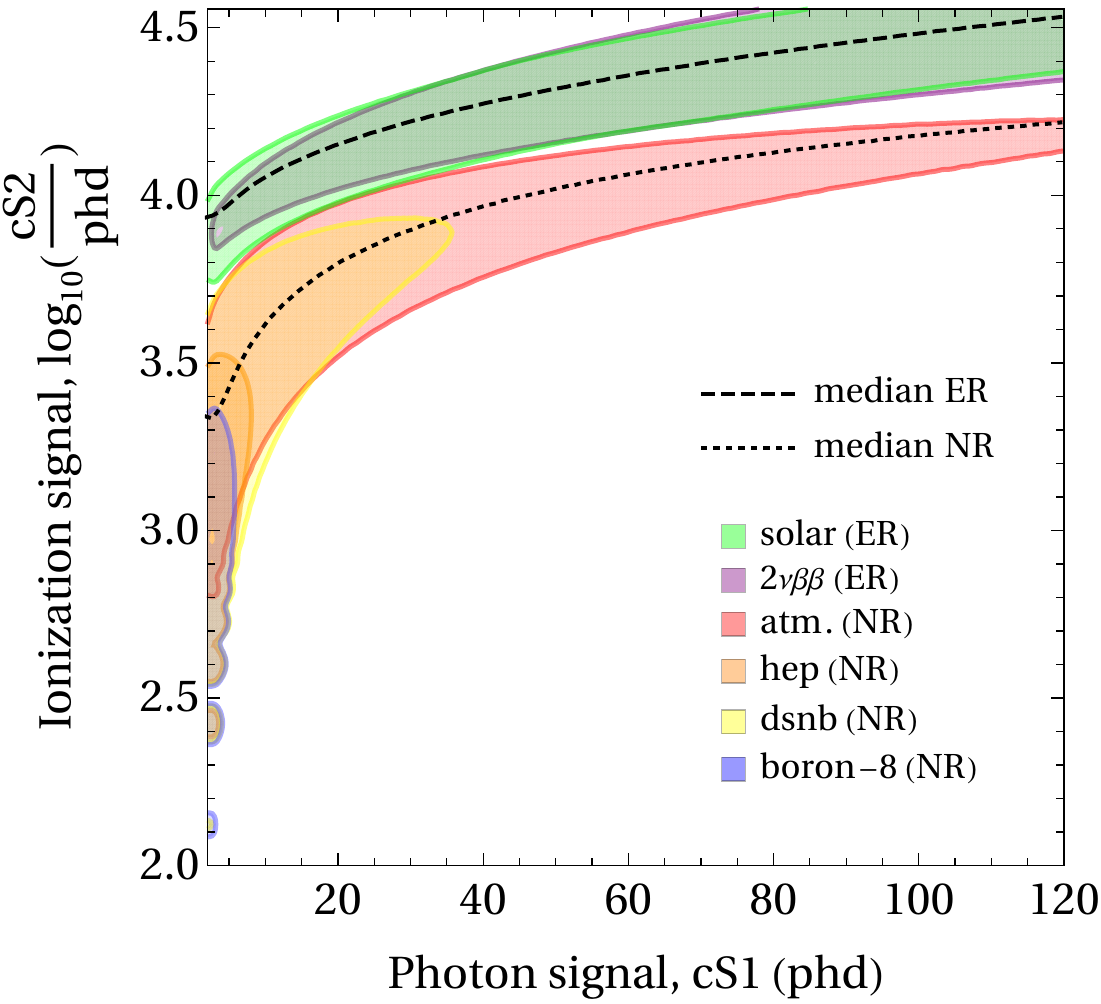}
    \includegraphics[width=0.40\linewidth]{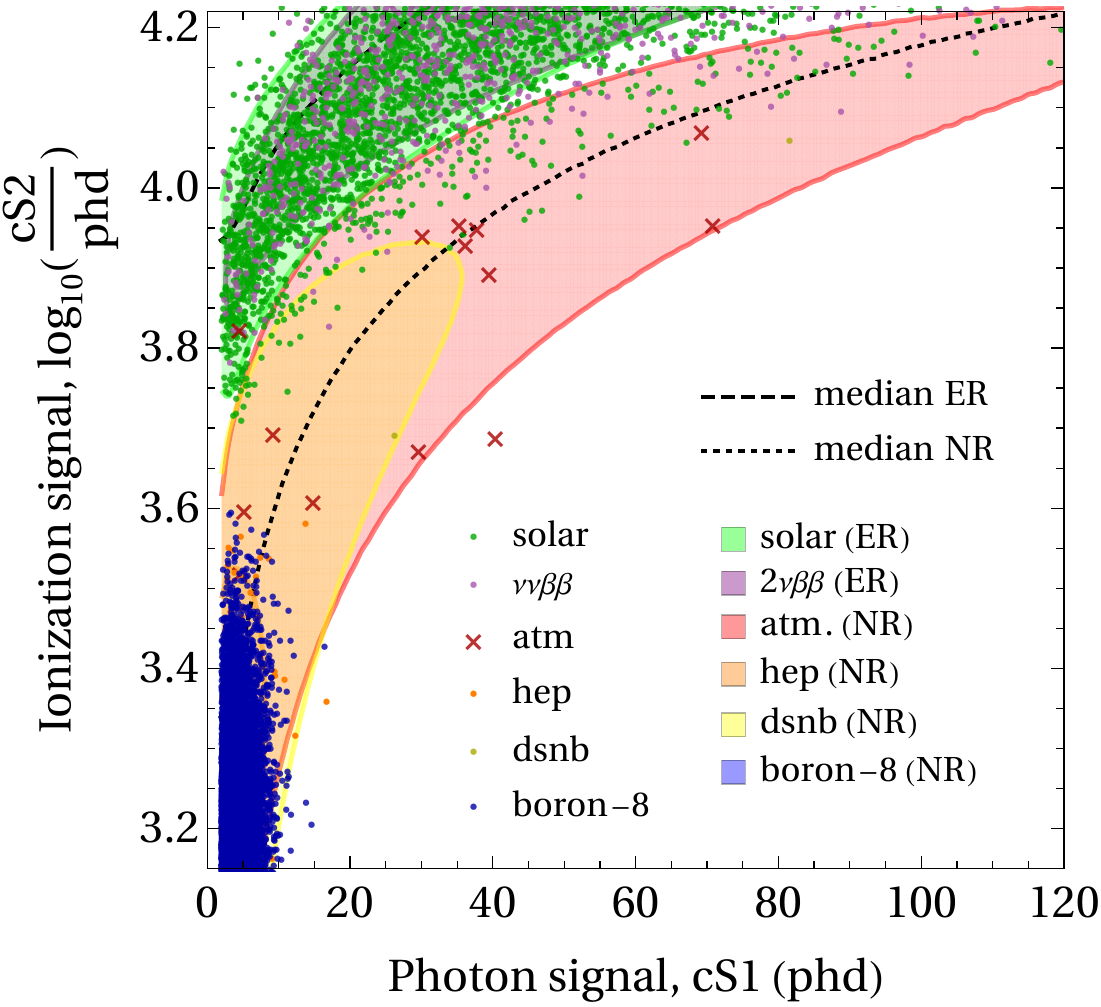}
    \caption{Left: Regions which contain 90\% of the events due to the specified source. The dashed (dotted) line shows the median of the nuclear (electronic) recoil band. Right: Same as right panel, except now assuming an exposure of 0.2 kilotonne-years, and zooming in on the vertical axis. Points represent simulated events from the indicated flux components.}
    \label{fig:nuS1S2}
\end{figure*}

We model a future Generation-3 xenon detector as a scaled-up version of the LZ detector, with dimensions scaled to obtain a fiducial region of 100~tonnes. The detector is modeled using the NESTv2 code which simulates the detailed micro-physics of the quanta production, recombination and final signal detection for electronic and nuclear recoil events in xenon~\cite{szydagis_m_2019_3357973}. We use a series of different detector configurations to investigate the effect of the different detector parameters on the results of the analysis. The values for the parameters are chosen between two values: a baseline value, where the parameter is similar to that already achieved in XENON1T~\cite{Aprile:2019bbb,Aprile:2019dme}, and an enhanced value that we deem achievable based on ongoing work in the community. A summary of these configurations is given in Table~\ref{tab:detPars}. In all configurations, we take the double-photoelectron detection probability to be 22\%, the number of PMTs to be 1200, and we require a 3-fold PMT coincidence for detection. Further details of the detector parameters, including the PMT properties and geometry, are taken from Refs.~\cite{Aprile:2017aty,Mount:2017qzi}. All analysis code, configuration files and results are publicly available for download~\cite{zenodo}.

\begin{table*}[htbp]
\caption{List of detector configurations and their corresponding parameters modelled in NEST. Note that $g_2$ is a derived parameter calculated from more fundamental detector parameters, see \cite{zenodo} for the full detector files used for this analysis.
\label{tab:detPars}
}
\begin{tabular}{ccccc}
\hline
Configuration & $g_1\,\,\,(\mathrm{phd}/\gamma)$ &  $g_2$ (phd/e) & drift field (V/cm) & electron lifetime ($\mu s$) \\
\hline
baseline       &  0.12 &  44 & 100  & 650\\
enhanced $g_1$ &  \textbf{0.3}  &  44 & 100  & 650\\
enhanced $g_2$ &  0.12 & \textbf{100} & 100  & 650\\
enhanced V     &  0.12 &  44 & \textbf{1000} & 650\\
enhanced $e$-lifetime &  0.12 &  44 & 100  & \textbf{5000}\\
all enhanced          &  0.3  & 100 & 1000 & 5000\\
\hline
\end{tabular}
\end{table*}

For comparison we also model a future Generation-3 argon detector. As argon detectors are able to achieve excellent electronic/nuclear recoil discrimination, no detailed detector simulation is required. We instead assume perfect discrimination above nuclear recoil energies of 25~keV, i.e. zero electronic recoil background in the region of interest. Proper modelling of the detector would be able to account for the roll-off of discrimination ability at low energies, allowing one to lower the threshold at a cost of efficiency. However, since the atmospheric rate is not strongly dependent on threshold, the small increase in signal would only have a correspondingly small effect on the present analysis. 

\subsection{Background components}

In this analysis we only include intrinsic backgrounds to an atmospheric neutrino search in xenon, assuming all other backgrounds are subdominant. This seems realistic in light of the current state of the art, with only mild extrapolation needed to a Generation-3 detector. The irreducible background consist of: electronic recoils from solar $pp$ and ${}^7$Be neutrinos, nuclear recoils from solar ${}^8$B and $hep$ neutrinos, the diffuse supernova neutrino background (dsnb), and the $\nu\nu\beta\beta$ decay of $^{136}$Xe. The $\nu\nu\beta\beta$ background could be suppressed through depletion of $^{136}$Xe, as explored in~\cite{Newstead:2018muu}. Here however we assume no depletion, as $^{136}$Xe is not the dominant background and will likely be desirable for a $0\nu\beta\beta$ search.

In this work, we use calculations of the solar neutrino electronic recoil rate from Ref.~\cite{Chen:2016eab}, which account for a $\sim 23\%$ suppression of the rate due to atomic binding effects. Additionally, we account for a $\sim 9\%$ reduction of the charge yield for L-shell electron recoils, as recently observed in electron-capture calibrations of the XELDA detector~\cite{dylanTalk}. This has the effect of widening the solar neutrino electronic recoil band and thus slightly increases the number of electronic recoil background events in the nuclear recoil signal band.

\subsection{Likelihood analysis}

To evaluate the future potential for discovery and measurement of the atmospheric neutrino flux, we perform a binned likelihood analysis on representative (Asimov) data sets~\cite{Cowan:2010js}, simulated with various detector exposures. To generate these data sets, we first perform a Monte Carlo simulation for each detector configuration, with $10^8$ events for each source of neutrinos. To investigate the effect of retaining position information in our likelihood, we obtained simulated distributions of events in two spaces: \{cS1, cS2\} and \{S1,S2,$r$,$z$\}, where {\it c} in cS1 refers to the S1 signal after correcting for position-dependent effects (as performed by NEST), and $r$ and $z$ refer to the radius and depth of the event in the detector. 

The distributions obtained for the `all enhanced' detector configuration are shown in the left panel of Figure~\ref{fig:nuS1S2} at the 90\% confidence level. The separation of signal and background regions can be deceiving since the expected rate for the solar components are orders of magnitude greater than for the atmospheric rate. To visualize the leakage of background events into the expected atmospheric background region, we therefore show a sample exposure of 0.2~kilotonne-years (kty) in the right panel of Figure~\ref{fig:nuS1S2}. This sample exposure highlights the futility of trying to define a background free region for an atmospheric neutrino search and why we must rely on statistical discrimination in the \{cS1, cS2\} plane. 

\begin{figure*}[ht!]
\includegraphics[width=0.75\columnwidth]{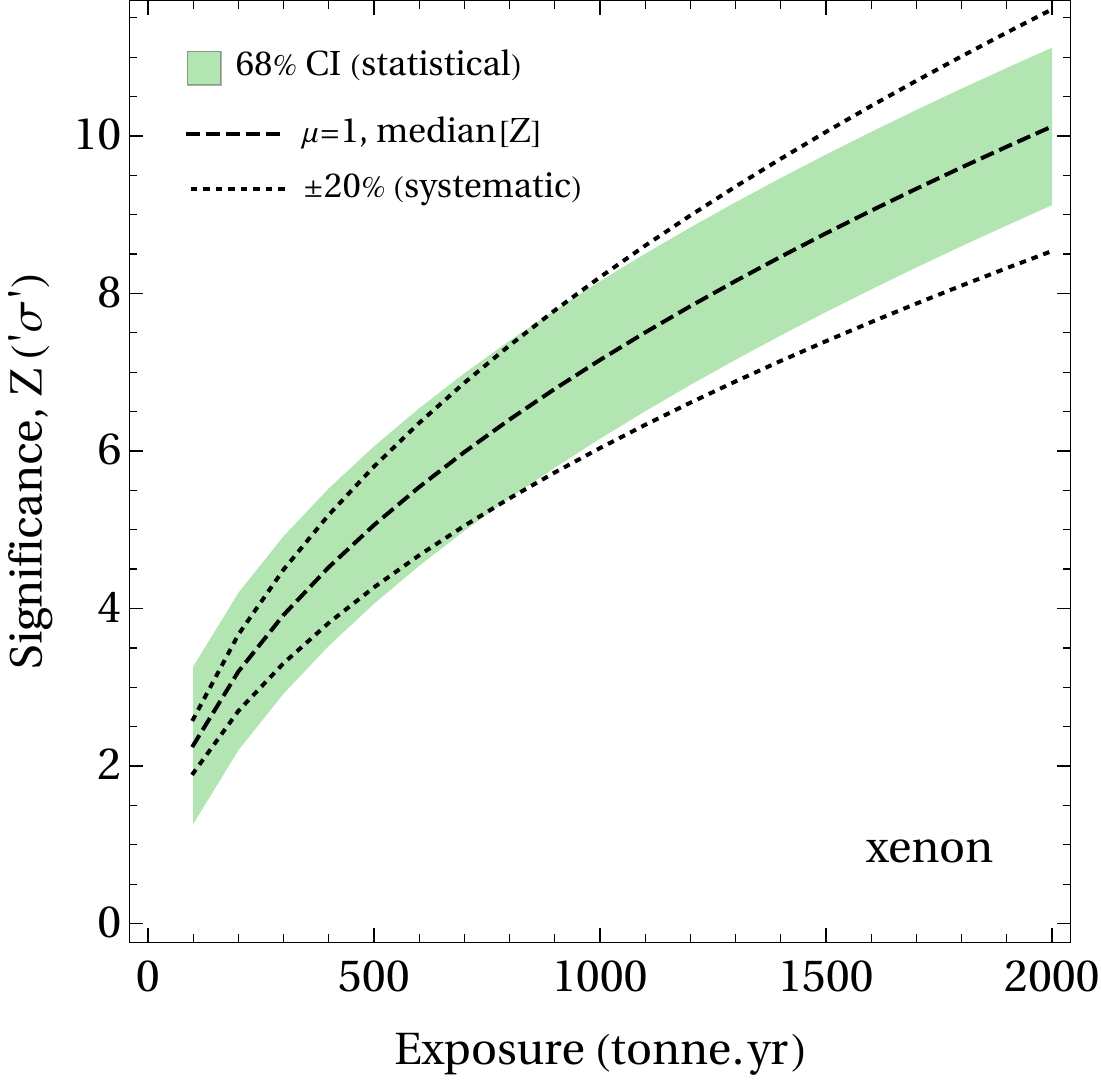}
\includegraphics[width=0.75\columnwidth]{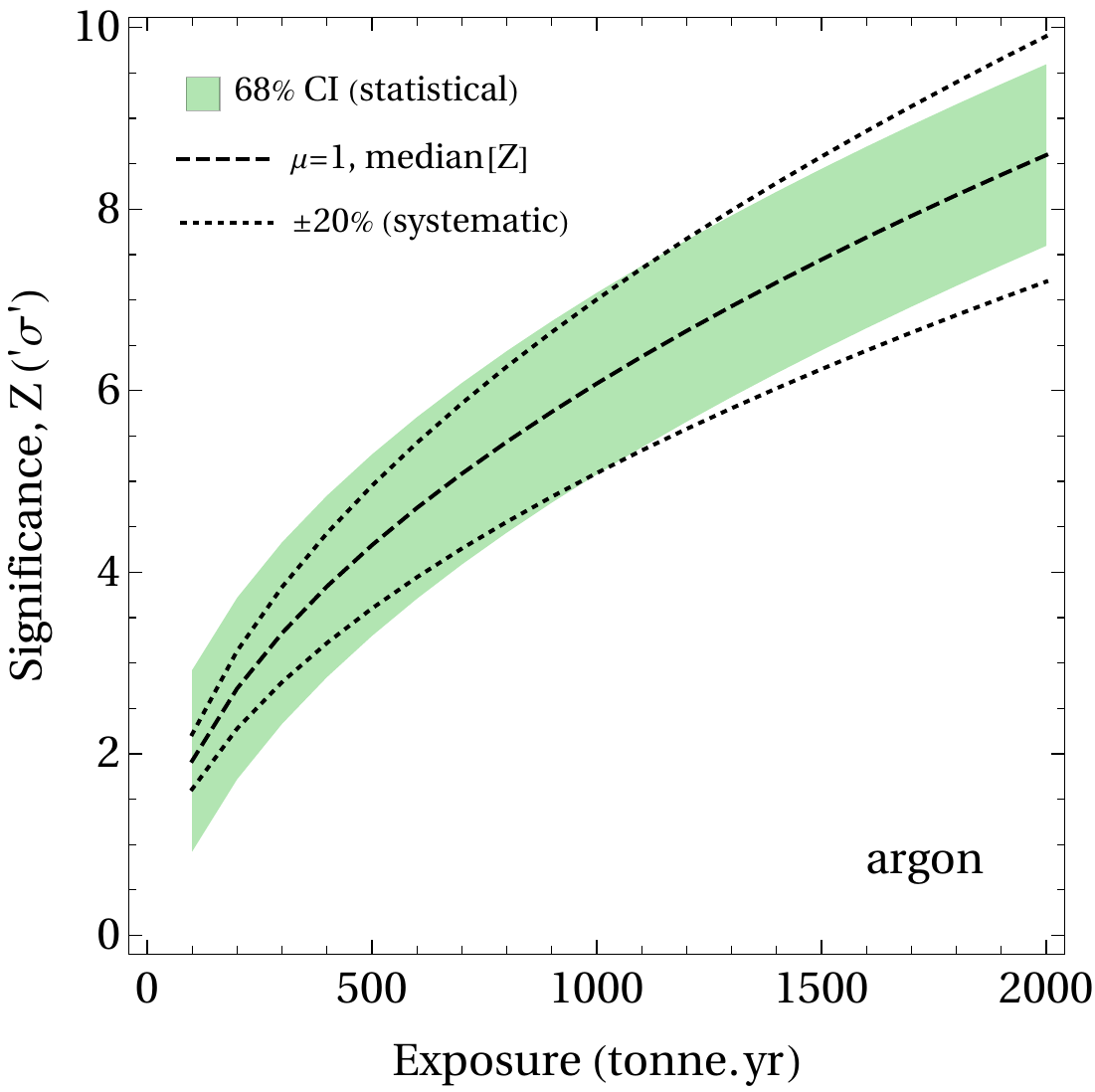}
\caption{The median significance (dashed) and 1-$\sigma$ confidence interval, CI, band (green) obtained for xenon (left) and argon (right) detectors as a function of the exposure. The effect of variations in the flux normalization by $\pm 20\%$ on the median significance is also shown (dotted). }
\label{fig:significance}
\end{figure*}

The analysis is performed in a region-of-interest defined by: $2 \leq \mathrm{cS1} \leq 120$ and $ 2 \leq \mathrm{log}_{10}(\mathrm{cS2}) \leq 4.56$. Extending this range does not improve our results statistically, so this range is chosen to reduce the computational burden of the analysis by allowing us to focus our simulation on the regions where our signal events are expected. These regions are divided into $N_{\mathrm{S1}}=120$ and $N_{\mathrm{S2}}=120$~bins, and the event positions are divided into $N_r=3$ and $N_z=5$~bins. The Poisson likelihood function is thus,
\be
\mathcal{L}({\bm n}|{\bm \lambda}(\mu'))=\sum_{i=1}^{N_{S1}}\sum_{j=1}^{N_{S2}}\sum_{k=1}^{N_{r}}\sum_{l=1}^{N_{z}} \mathrm{Poisson}(n_{i,j,k,l}|\lambda_{i,j,k,l}( \mu'))
\ee
where $\mathrm{Poisson}(n_{i,j,k,l}|\lambda_{i,j,k,l}(\mu'))$ is the Poisson probability of observing $n_{i,j,k,l}$ events in the ${i,j,k,l}$ bin, given an expected (mean) number of events, 
\be
\lambda_{i,j,k,l}(\mu')=b_{i,j,k,l}+\mu' s_{i,j,k,l},
\ee
for a given signal strength, $\mu'$, expected background, $b$, and signal, $s$. 

To calculate the expected statistical significance of discovery for a given exposure and assumed signal strength we use the test statistic
\be
\label{eq:q0}
q_{0,A}(\mu') = -2 \mathrm{ln}\frac{\mathcal{L}(\mu=0\vert {\bm\lambda}( \mu'))}{\mathcal{L}(\hat\mu=1\vert {\bm\lambda}(\mu'))}, 
\ee
where the expected significance is given by $\sqrt{q_{0,A}}$. We additionally calculate the 90\% confidence interval on the signal strength parameter using the test statistic, 
\be
t_\mu(\mu') = -2 \mathrm{ln}\frac{\mathcal{L}(\mu=0\vert {\bm\lambda(\mu')}}{\mathcal{L}(\hat\mu=1\vert{\bm\lambda}(\mu'))}. 
\ee

\section{Results}

To explore the effect of using different detector parameters (as given in Table~\ref{tab:detPars}) we perform our analysis six times with different configurations. First using the baseline values and then the enhanced values one at a time, and then all together. The expected significance of the atmospheric neutrino signal after 1~kty is given for each of the configurations in Table~\ref{tab:results}. While there may be significant correlations between pairs of the parameters, these results clearly show that the parameter with the single largest effect is the mean electron lifetime. This is because most of the parameters are independent of, or scale with, the detector size. This is not the case for the electron lifetime which, if left at 650~$\mu$s, results in far fewer electrons making it to the liquid surface. Electron lifetimes longer than 5~seconds where tested and found to not further improve the detector performance.

\begin{table}[htbp]
  \caption{The expected significance (Z) of the atmospheric neutrino signal for each of the detector configurations with a 1~kty exposure, with and without position information.}
    \centering
    \begin{tabular}{ccc}
    \hline
      Enhanced  &  significance Z & significance Z \\
      parameter &  (corrected S1 \& S2) &  (position bins) \\
 \hline
        N/A               & 1.2 & 1.5\\
        $g_1$             & 1.6 & 2.0\\
        $g_2$             & 1.3 & 1.5\\
        drift field       & 1.2 & 1.6\\
        electron lifetime & 5.7 & 6.0\\
        all               & 7.0 & 7.2\\
    \hline
    \end{tabular}
    \label{tab:results}
\end{table}

The detector configuration with all parameters enhanced experienced an improvement beyond the sum of the individual improvements, indicating that the enhancements are acting synergistically. This can be explained by the observation that increasing $g_2$ alone has a negligible improvement on the sensitivity. One would expect that an increase in the gain of the S2 signal would improve electronic/nuclear recoil discrimination by reducing statistical fluctuations in the S2 signal (and also improve position reconstruction). However, if the electron lifetime is small, then there is no signal for $g_2$ to amplify, and no improvement is observed. We find that including position information in the likelihood increases the significance for all detector configurations, with the worse performing configurations seeing the most improvement.

\begin{figure}[htbp]
\includegraphics[width=0.9\columnwidth]{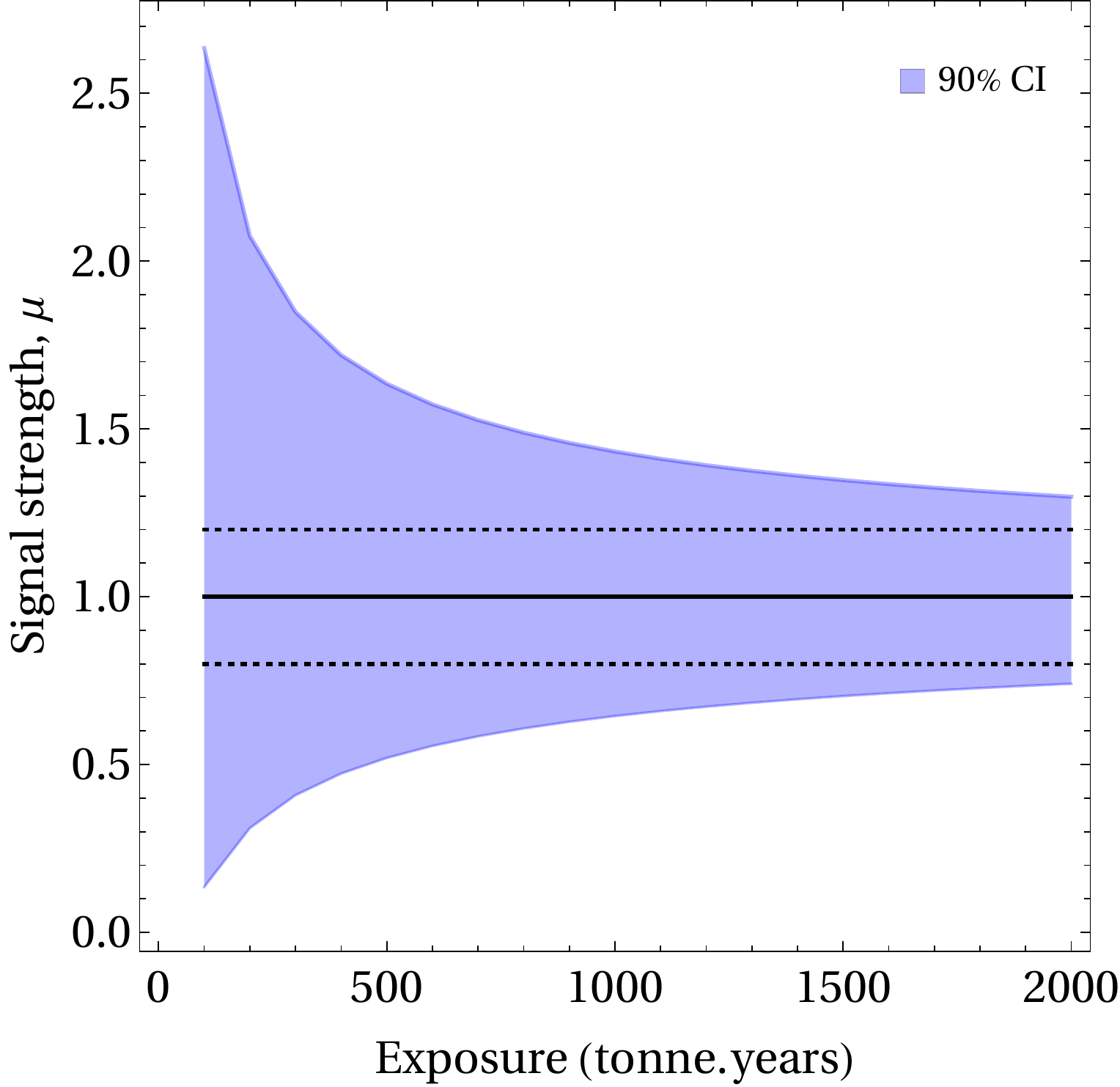}
\caption{The 90\% confidence interval of the measured atmospheric flux normalization for xenon (blue). The 20\% systematic uncertainty in the atmospheric flux is shown for comparison (dotted).}
\label{fig:CI}
\end{figure}

To show the effect of the systematic uncertainty on the projected significance, we perform our significance calculation varying the flux normalization by $\pm 20\%$, i.e. using Eq.~\ref{eq:q0} with values of $\mu'={0.8,1,1.2}$. The results of these calculations for both xenon and argon detectors are shown in Fig.~\ref{fig:significance}. At an exposure of 1~kty, the systematic uncertainty is approximately $\pm 1\sigma$, which is of the same magnitude as the statistical uncertainty. Figure~\ref{fig:CI} shows the corresponding confidence interval obtained for a measurement of the atmospheric flux normalization with xenon.

\section{Discussion and Conclusion} 

The standard paradigm of dark matter direct detection experiments is to operate as close to the no-background regime as possible. It has long been known that this paradigm would come to an end when the irreducible nuclear recoil backgrounds from neutrinos dominate and sensitivities reach the so called `neutrino floor'.  For such nuclear recoil searches, the ability to discriminate and reject the electron recoil background is crucial. In this paper we find that even before reaching the atmospheric background, realistic detectors will observe many solar neutrinos events leaking into the nuclear recoil band (as can be clearly seen in Fig.~\ref{fig:nuS1S2}). We have shown that the standard practice of using a 2D likelihood in S1 and S2 provides enough discrimination power to make the first measurement of atmospheric neutrinos. Critically, we have identified that a good electron lifetime is necessary for such a measurement.

We have discussed the prospects for extracting the atmospheric neutrino flux at multi-ton scale xenon experiments. Using the best estimates for the low-energy atmospheric neutrino flux, and assuming pure Standard Model physics, we find that an exposure of at least 0.5~kty (0.7~kty) is required to obtain a $5\sigma$ detection in xenon (argon). This is well within reach of a future Generation-3 detector technology.

The detectors discussed will not only be important for atmospheric neutrinos, but they will provide a new window into solar and supernova neutrinos as well. In addition to focusing on detector specifications, we highlight the need to better understand the systematic uncertainty of the low-energy atmospheric neutrino flux. While a Generation-3 xenon detector could be used to make the first measurement of this flux, with the current technology and exposure of less than 2~kty, such a measurement would not be able to measure the flux to better than 20\% precision. This does not account for non-standard neutrino interactions, which may alter the flux relative to the predictions presented in this paper~\cite{Gonzalez-Garcia:2018dep,Boehm:2018sux}, and may be studied with Generation-3 experiments~\cite{Dutta:2020che}.

\section{Acknowledgements}

We thank Amanda Depoian for useful comments on this manuscript. The work of LES is supported by DOE grant de-sc0010813 and that of RFL and JLN by NSF grant PHY-1719271. JLN is supported in part by the Australian Research Council. JLN thanks The Mitchell Institute for their hospitality while a portion of this work was completed. 

\bibliographystyle{apsrev4-1}
\bibliography{bibliography.bib}

\begin{thebibliography}{33}%
\makeatletter
\providecommand \@ifxundefined [1]{%
 \@ifx{#1\undefined}
}%
\providecommand \@ifnum [1]{%
 \ifnum #1\expandafter \@firstoftwo
 \else \expandafter \@secondoftwo
 \fi
}%
\providecommand \@ifx [1]{%
 \ifx #1\expandafter \@firstoftwo
 \else \expandafter \@secondoftwo
 \fi
}%
\providecommand \natexlab [1]{#1}%
\providecommand \enquote  [1]{``#1''}%
\providecommand \bibnamefont  [1]{#1}%
\providecommand \bibfnamefont [1]{#1}%
\providecommand \citenamefont [1]{#1}%
\providecommand \href@noop [0]{\@secondoftwo}%
\providecommand \href [0]{\begingroup \@sanitize@url \@href}%
\providecommand \@href[1]{\@@startlink{#1}\@@href}%
\providecommand \@@href[1]{\endgroup#1\@@endlink}%
\providecommand \@sanitize@url [0]{\catcode `\\12\catcode `\$12\catcode
  `\&12\catcode `\#12\catcode `\^12\catcode `\_12\catcode `\%12\relax}%
\providecommand \@@startlink[1]{}%
\providecommand \@@endlink[0]{}%
\providecommand \url  [0]{\begingroup\@sanitize@url \@url }%
\providecommand \@url [1]{\endgroup\@href {#1}{\urlprefix }}%
\providecommand \urlprefix  [0]{URL }%
\providecommand \Eprint [0]{\href }%
\providecommand \doibase [0]{http://dx.doi.org/}%
\providecommand \selectlanguage [0]{\@gobble}%
\providecommand \bibinfo  [0]{\@secondoftwo}%
\providecommand \bibfield  [0]{\@secondoftwo}%
\providecommand \translation [1]{[#1]}%
\providecommand \BibitemOpen [0]{}%
\providecommand \bibitemStop [0]{}%
\providecommand \bibitemNoStop [0]{.\EOS\space}%
\providecommand \EOS [0]{\spacefactor3000\relax}%
\providecommand \BibitemShut  [1]{\csname bibitem#1\endcsname}%
\let\auto@bib@innerbib\@empty
\bibitem [{\citenamefont {Billard}\ \emph {et~al.}(2014)\citenamefont
  {Billard}, \citenamefont {Strigari},\ and\ \citenamefont
  {Figueroa-Feliciano}}]{Billard:2013qya}%
  \BibitemOpen
  \bibfield  {author} {\bibinfo {author} {\bibfnamefont {J.}~\bibnamefont
  {Billard}}, \bibinfo {author} {\bibfnamefont {L.}~\bibnamefont {Strigari}}, \
  and\ \bibinfo {author} {\bibfnamefont {E.}~\bibnamefont
  {Figueroa-Feliciano}},\ }\href {\doibase 10.1103/PhysRevD.89.023524}
  {\bibfield  {journal} {\bibinfo  {journal} {Phys. Rev.}\ }\textbf {\bibinfo
  {volume} {D89}},\ \bibinfo {pages} {023524} (\bibinfo {year} {2014})},\
  \Eprint {http://arxiv.org/abs/1307.5458} {arXiv:1307.5458 [hep-ph]}
  \BibitemShut {NoStop}%
\bibitem [{\citenamefont {Billard}\ \emph {et~al.}(2015)\citenamefont
  {Billard}, \citenamefont {Strigari},\ and\ \citenamefont
  {Figueroa-Feliciano}}]{Billard:2014yka}%
  \BibitemOpen
  \bibfield  {author} {\bibinfo {author} {\bibfnamefont {J.}~\bibnamefont
  {Billard}}, \bibinfo {author} {\bibfnamefont {L.~E.}\ \bibnamefont
  {Strigari}}, \ and\ \bibinfo {author} {\bibfnamefont {E.}~\bibnamefont
  {Figueroa-Feliciano}},\ }\href {\doibase 10.1103/PhysRevD.91.095023}
  {\bibfield  {journal} {\bibinfo  {journal} {Phys. Rev.}\ }\textbf {\bibinfo
  {volume} {D91}},\ \bibinfo {pages} {095023} (\bibinfo {year} {2015})},\
  \Eprint {http://arxiv.org/abs/1409.0050} {arXiv:1409.0050 [astro-ph.CO]}
  \BibitemShut {NoStop}%
\bibitem [{\citenamefont {Lang}\ \emph {et~al.}(2016)\citenamefont {Lang},
  \citenamefont {McCabe}, \citenamefont {Reichard}, \citenamefont {Selvi},\
  and\ \citenamefont {Tamborra}}]{Lang:2016zhv}%
  \BibitemOpen
  \bibfield  {author} {\bibinfo {author} {\bibfnamefont {R.~F.}\ \bibnamefont
  {Lang}}, \bibinfo {author} {\bibfnamefont {C.}~\bibnamefont {McCabe}},
  \bibinfo {author} {\bibfnamefont {S.}~\bibnamefont {Reichard}}, \bibinfo
  {author} {\bibfnamefont {M.}~\bibnamefont {Selvi}}, \ and\ \bibinfo {author}
  {\bibfnamefont {I.}~\bibnamefont {Tamborra}},\ }\href {\doibase
  10.1103/PhysRevD.94.103009} {\bibfield  {journal} {\bibinfo  {journal} {Phys.
  Rev.}\ }\textbf {\bibinfo {volume} {D94}},\ \bibinfo {pages} {103009}
  (\bibinfo {year} {2016})},\ \Eprint {http://arxiv.org/abs/1606.09243}
  {arXiv:1606.09243 [astro-ph.HE]} \BibitemShut {NoStop}%
\bibitem [{\citenamefont {Strigari}(2009)}]{Strigari:2009bq}%
  \BibitemOpen
  \bibfield  {author} {\bibinfo {author} {\bibfnamefont {L.~E.}\ \bibnamefont
  {Strigari}},\ }\href {\doibase 10.1088/1367-2630/11/10/105011} {\bibfield
  {journal} {\bibinfo  {journal} {New J. Phys.}\ }\textbf {\bibinfo {volume}
  {11}},\ \bibinfo {pages} {105011} (\bibinfo {year} {2009})},\ \Eprint
  {http://arxiv.org/abs/0903.3630} {arXiv:0903.3630 [astro-ph.CO]} \BibitemShut
  {NoStop}%
\bibitem [{\citenamefont {Dutta}\ and\ \citenamefont
  {Strigari}(2019)}]{Dutta:2019oaj}%
  \BibitemOpen
  \bibfield  {author} {\bibinfo {author} {\bibfnamefont {B.}~\bibnamefont
  {Dutta}}\ and\ \bibinfo {author} {\bibfnamefont {L.~E.}\ \bibnamefont
  {Strigari}},\ }\href {\doibase 10.1146/annurev-nucl-101918-023450} {\bibfield
   {journal} {\bibinfo  {journal} {Ann. Rev. Nucl. Part. Sci.}\ }\textbf
  {\bibinfo {volume} {69}},\ \bibinfo {pages} {137} (\bibinfo {year} {2019})},\
  \Eprint {http://arxiv.org/abs/1901.08876} {arXiv:1901.08876 [hep-ph]}
  \BibitemShut {NoStop}%
\bibitem [{\citenamefont {Aalbers}\ \emph {et~al.}(2016)\citenamefont {Aalbers}
  \emph {et~al.}}]{Aalbers:2016jon}%
  \BibitemOpen
  \bibfield  {author} {\bibinfo {author} {\bibfnamefont {J.}~\bibnamefont
  {Aalbers}} \emph {et~al.} (\bibinfo {collaboration} {DARWIN}),\ }\href
  {\doibase 10.1088/1475-7516/2016/11/017} {\bibfield  {journal} {\bibinfo
  {journal} {JCAP}\ }\textbf {\bibinfo {volume} {1611}},\ \bibinfo {pages}
  {017} (\bibinfo {year} {2016})},\ \Eprint {http://arxiv.org/abs/1606.07001}
  {arXiv:1606.07001 [astro-ph.IM]} \BibitemShut {NoStop}%
\bibitem [{\citenamefont {Schumann}\ \emph {et~al.}(2015)\citenamefont
  {Schumann}, \citenamefont {Baudis}, \citenamefont {Bütikofer}, \citenamefont
  {Kish},\ and\ \citenamefont {Selvi}}]{Schumann:2015cpa}%
  \BibitemOpen
  \bibfield  {author} {\bibinfo {author} {\bibfnamefont {M.}~\bibnamefont
  {Schumann}}, \bibinfo {author} {\bibfnamefont {L.}~\bibnamefont {Baudis}},
  \bibinfo {author} {\bibfnamefont {L.}~\bibnamefont {Bütikofer}}, \bibinfo
  {author} {\bibfnamefont {A.}~\bibnamefont {Kish}}, \ and\ \bibinfo {author}
  {\bibfnamefont {M.}~\bibnamefont {Selvi}},\ }\href {\doibase
  10.1088/1475-7516/2015/10/016} {\bibfield  {journal} {\bibinfo  {journal}
  {JCAP}\ }\textbf {\bibinfo {volume} {1510}},\ \bibinfo {pages} {016}
  (\bibinfo {year} {2015})},\ \Eprint {http://arxiv.org/abs/1506.08309}
  {arXiv:1506.08309 [physics.ins-det]} \BibitemShut {NoStop}%
\bibitem [{\citenamefont {Dent}\ \emph {et~al.}(2017)\citenamefont {Dent},
  \citenamefont {Dutta}, \citenamefont {Newstead},\ and\ \citenamefont
  {Strigari}}]{Dent:2016wor}%
  \BibitemOpen
  \bibfield  {author} {\bibinfo {author} {\bibfnamefont {J.~B.}\ \bibnamefont
  {Dent}}, \bibinfo {author} {\bibfnamefont {B.}~\bibnamefont {Dutta}},
  \bibinfo {author} {\bibfnamefont {J.~L.}\ \bibnamefont {Newstead}}, \ and\
  \bibinfo {author} {\bibfnamefont {L.~E.}\ \bibnamefont {Strigari}},\ }\href
  {\doibase 10.1103/PhysRevD.95.051701} {\bibfield  {journal} {\bibinfo
  {journal} {Phys. Rev.}\ }\textbf {\bibinfo {volume} {D95}},\ \bibinfo {pages}
  {051701} (\bibinfo {year} {2017})},\ \Eprint
  {http://arxiv.org/abs/1607.01468} {arXiv:1607.01468 [hep-ph]} \BibitemShut
  {NoStop}%
\bibitem [{\citenamefont {Dent}\ \emph {et~al.}(2016)\citenamefont {Dent},
  \citenamefont {Dutta}, \citenamefont {Newstead},\ and\ \citenamefont
  {Strigari}}]{Dent:2016iht}%
  \BibitemOpen
  \bibfield  {author} {\bibinfo {author} {\bibfnamefont {J.~B.}\ \bibnamefont
  {Dent}}, \bibinfo {author} {\bibfnamefont {B.}~\bibnamefont {Dutta}},
  \bibinfo {author} {\bibfnamefont {J.~L.}\ \bibnamefont {Newstead}}, \ and\
  \bibinfo {author} {\bibfnamefont {L.~E.}\ \bibnamefont {Strigari}},\ }\href
  {\doibase 10.1103/PhysRevD.93.075018} {\bibfield  {journal} {\bibinfo
  {journal} {Phys. Rev.}\ }\textbf {\bibinfo {volume} {D93}},\ \bibinfo {pages}
  {075018} (\bibinfo {year} {2016})},\ \Eprint
  {http://arxiv.org/abs/1602.05300} {arXiv:1602.05300 [hep-ph]} \BibitemShut
  {NoStop}%
\bibitem [{\citenamefont {O'Hare}(2020)}]{OHare:2020lva}%
  \BibitemOpen
  \bibfield  {author} {\bibinfo {author} {\bibfnamefont {C.~A.~J.}\
  \bibnamefont {O'Hare}},\ }\href@noop {} {\  (\bibinfo {year} {2020})},\
  \Eprint {http://arxiv.org/abs/2002.07499} {arXiv:2002.07499 [astro-ph.CO]}
  \BibitemShut {NoStop}%
\bibitem [{\citenamefont {Newstead}\ \emph {et~al.}(2019)\citenamefont
  {Newstead}, \citenamefont {Strigari},\ and\ \citenamefont
  {Lang}}]{Newstead:2018muu}%
  \BibitemOpen
  \bibfield  {author} {\bibinfo {author} {\bibfnamefont {J.~L.}\ \bibnamefont
  {Newstead}}, \bibinfo {author} {\bibfnamefont {L.~E.}\ \bibnamefont
  {Strigari}}, \ and\ \bibinfo {author} {\bibfnamefont {R.~F.}\ \bibnamefont
  {Lang}},\ }\href {\doibase 10.1103/PhysRevD.99.043006} {\bibfield  {journal}
  {\bibinfo  {journal} {Phys. Rev.}\ }\textbf {\bibinfo {volume} {D99}},\
  \bibinfo {pages} {043006} (\bibinfo {year} {2019})},\ \Eprint
  {http://arxiv.org/abs/1807.07169} {arXiv:1807.07169 [astro-ph.SR]}
  \BibitemShut {NoStop}%
\bibitem [{\citenamefont {Szydagis}\ \emph {et~al.}(2011)\citenamefont
  {Szydagis}, \citenamefont {Barry}, \citenamefont {Kazkaz}, \citenamefont
  {Mock}, \citenamefont {Stolp}, \citenamefont {Sweany}, \citenamefont
  {Tripathi}, \citenamefont {Uvarov}, \citenamefont {Walsh},\ and\
  \citenamefont {Woods}}]{Szydagis:2011tk}%
  \BibitemOpen
  \bibfield  {author} {\bibinfo {author} {\bibfnamefont {M.}~\bibnamefont
  {Szydagis}}, \bibinfo {author} {\bibfnamefont {N.}~\bibnamefont {Barry}},
  \bibinfo {author} {\bibfnamefont {K.}~\bibnamefont {Kazkaz}}, \bibinfo
  {author} {\bibfnamefont {J.}~\bibnamefont {Mock}}, \bibinfo {author}
  {\bibfnamefont {D.}~\bibnamefont {Stolp}}, \bibinfo {author} {\bibfnamefont
  {M.}~\bibnamefont {Sweany}}, \bibinfo {author} {\bibfnamefont
  {M.}~\bibnamefont {Tripathi}}, \bibinfo {author} {\bibfnamefont
  {S.}~\bibnamefont {Uvarov}}, \bibinfo {author} {\bibfnamefont
  {N.}~\bibnamefont {Walsh}}, \ and\ \bibinfo {author} {\bibfnamefont
  {M.}~\bibnamefont {Woods}},\ }\href {\doibase 10.1088/1748-0221/6/10/P10002}
  {\bibfield  {journal} {\bibinfo  {journal} {JINST}\ }\textbf {\bibinfo
  {volume} {6}},\ \bibinfo {pages} {P10002} (\bibinfo {year} {2011})},\ \Eprint
  {http://arxiv.org/abs/1106.1613} {arXiv:1106.1613 [physics.ins-det]}
  \BibitemShut {NoStop}%
\bibitem [{\citenamefont {Szydagis}\ \emph {et~al.}(2019)\citenamefont
  {Szydagis}, \citenamefont {Balajthy}, \citenamefont {Brodsky}, \citenamefont
  {Cutter}, \citenamefont {Huang}, \citenamefont {Kozlova}, \citenamefont
  {Lenardo}, \citenamefont {Manalaysay}, \citenamefont {McKinsey},
  \citenamefont {Mooney}, \citenamefont {Mueller}, \citenamefont {Ni},
  \citenamefont {Rischbieter}, \citenamefont {Tripathi}, \citenamefont
  {Tunnell}, \citenamefont {Velan},\ and\ \citenamefont
  {Zhao}}]{szydagis_m_2019_3357973}%
  \BibitemOpen
  \bibfield  {author} {\bibinfo {author} {\bibfnamefont {M.}~\bibnamefont
  {Szydagis}}, \bibinfo {author} {\bibfnamefont {J.}~\bibnamefont {Balajthy}},
  \bibinfo {author} {\bibfnamefont {J.}~\bibnamefont {Brodsky}}, \bibinfo
  {author} {\bibfnamefont {J.}~\bibnamefont {Cutter}}, \bibinfo {author}
  {\bibfnamefont {J.}~\bibnamefont {Huang}}, \bibinfo {author} {\bibfnamefont
  {E.}~\bibnamefont {Kozlova}}, \bibinfo {author} {\bibfnamefont
  {B.}~\bibnamefont {Lenardo}}, \bibinfo {author} {\bibfnamefont
  {A.}~\bibnamefont {Manalaysay}}, \bibinfo {author} {\bibfnamefont
  {D.}~\bibnamefont {McKinsey}}, \bibinfo {author} {\bibfnamefont
  {M.}~\bibnamefont {Mooney}}, \bibinfo {author} {\bibfnamefont
  {J.}~\bibnamefont {Mueller}}, \bibinfo {author} {\bibfnamefont
  {K.}~\bibnamefont {Ni}}, \bibinfo {author} {\bibfnamefont {G.}~\bibnamefont
  {Rischbieter}}, \bibinfo {author} {\bibfnamefont {M.}~\bibnamefont
  {Tripathi}}, \bibinfo {author} {\bibfnamefont {C.}~\bibnamefont {Tunnell}},
  \bibinfo {author} {\bibfnamefont {V.}~\bibnamefont {Velan}}, \ and\ \bibinfo
  {author} {\bibfnamefont {Z.}~\bibnamefont {Zhao}},\ }\href {\doibase
  10.5281/zenodo.3357973} {\enquote {\bibinfo {title} {{NESTCollaboration/nest:
  New, flexible LXe NR yields and resolution model + G4 improvements + linear
  Noise + much more}},}\ } (\bibinfo {year} {2019})\BibitemShut {NoStop}%
\bibitem [{\citenamefont {Battistoni}\ \emph {et~al.}(2003)\citenamefont
  {Battistoni}, \citenamefont {Ferrari}, \citenamefont {Montaruli},\ and\
  \citenamefont {Sala}}]{Battistoni:2002ew}%
  \BibitemOpen
  \bibfield  {author} {\bibinfo {author} {\bibfnamefont {G.}~\bibnamefont
  {Battistoni}}, \bibinfo {author} {\bibfnamefont {A.}~\bibnamefont {Ferrari}},
  \bibinfo {author} {\bibfnamefont {T.}~\bibnamefont {Montaruli}}, \ and\
  \bibinfo {author} {\bibfnamefont {P.~R.}\ \bibnamefont {Sala}},\ }\href
  {\doibase 10.1016/S0927-6505(02)00246-3} {\bibfield  {journal} {\bibinfo
  {journal} {Astropart. Phys.}\ }\textbf {\bibinfo {volume} {19}},\ \bibinfo
  {pages} {269} (\bibinfo {year} {2003})},\ \bibinfo {note} {[Erratum:
  Astropart. Phys.19,291(2003)]},\ \Eprint
  {http://arxiv.org/abs/hep-ph/0207035} {arXiv:hep-ph/0207035 [hep-ph]}
  \BibitemShut {NoStop}%
\bibitem [{\citenamefont {Honda}\ \emph {et~al.}(2011)\citenamefont {Honda},
  \citenamefont {Kajita}, \citenamefont {Kasahara},\ and\ \citenamefont
  {Midorikawa}}]{Honda:2011nf}%
  \BibitemOpen
  \bibfield  {author} {\bibinfo {author} {\bibfnamefont {M.}~\bibnamefont
  {Honda}}, \bibinfo {author} {\bibfnamefont {T.}~\bibnamefont {Kajita}},
  \bibinfo {author} {\bibfnamefont {K.}~\bibnamefont {Kasahara}}, \ and\
  \bibinfo {author} {\bibfnamefont {S.}~\bibnamefont {Midorikawa}},\ }\href
  {\doibase 10.1103/PhysRevD.83.123001} {\bibfield  {journal} {\bibinfo
  {journal} {Phys. Rev.}\ }\textbf {\bibinfo {volume} {D83}},\ \bibinfo {pages}
  {123001} (\bibinfo {year} {2011})},\ \Eprint {http://arxiv.org/abs/1102.2688}
  {arXiv:1102.2688 [astro-ph.HE]} \BibitemShut {NoStop}%
\bibitem [{\citenamefont {Richard}\ \emph {et~al.}(2016)\citenamefont {Richard}
  \emph {et~al.}}]{Richard:2015aua}%
  \BibitemOpen
  \bibfield  {author} {\bibinfo {author} {\bibfnamefont {E.}~\bibnamefont
  {Richard}} \emph {et~al.} (\bibinfo {collaboration} {Super-Kamiokande}),\
  }\href {\doibase 10.1103/PhysRevD.94.052001} {\bibfield  {journal} {\bibinfo
  {journal} {Phys. Rev.}\ }\textbf {\bibinfo {volume} {D94}},\ \bibinfo {pages}
  {052001} (\bibinfo {year} {2016})},\ \Eprint
  {http://arxiv.org/abs/1510.08127} {arXiv:1510.08127 [hep-ex]} \BibitemShut
  {NoStop}%
\bibitem [{\citenamefont {Helm}(1956)}]{Helm:1956zz}%
  \BibitemOpen
  \bibfield  {author} {\bibinfo {author} {\bibfnamefont {R.~H.}\ \bibnamefont
  {Helm}},\ }\href {\doibase 10.1103/PhysRev.104.1466} {\bibfield  {journal}
  {\bibinfo  {journal} {Phys. Rev.}\ }\textbf {\bibinfo {volume} {104}},\
  \bibinfo {pages} {1466} (\bibinfo {year} {1956})}\BibitemShut {NoStop}%
\bibitem [{\citenamefont {An}\ \emph {et~al.}(2016)\citenamefont {An} \emph
  {et~al.}}]{An:2015jdp}%
  \BibitemOpen
  \bibfield  {author} {\bibinfo {author} {\bibfnamefont {F.}~\bibnamefont {An}}
  \emph {et~al.} (\bibinfo {collaboration} {JUNO}),\ }\href {\doibase
  10.1088/0954-3899/43/3/030401} {\bibfield  {journal} {\bibinfo  {journal} {J.
  Phys.}\ }\textbf {\bibinfo {volume} {G43}},\ \bibinfo {pages} {030401}
  (\bibinfo {year} {2016})},\ \Eprint {http://arxiv.org/abs/1507.05613}
  {arXiv:1507.05613 [physics.ins-det]} \BibitemShut {NoStop}%
\bibitem [{\citenamefont {Peres}\ and\ \citenamefont
  {Smirnov}(2009)}]{Peres:2009xe}%
  \BibitemOpen
  \bibfield  {author} {\bibinfo {author} {\bibfnamefont {O.~L.~G.}\
  \bibnamefont {Peres}}\ and\ \bibinfo {author} {\bibfnamefont {A.~{\relax
  Yu}.}\ \bibnamefont {Smirnov}},\ }\href {\doibase 10.1103/PhysRevD.79.113002}
  {\bibfield  {journal} {\bibinfo  {journal} {Phys. Rev.}\ }\textbf {\bibinfo
  {volume} {D79}},\ \bibinfo {pages} {113002} (\bibinfo {year} {2009})},\
  \Eprint {http://arxiv.org/abs/0903.5323} {arXiv:0903.5323 [hep-ph]}
  \BibitemShut {NoStop}%
\bibitem [{\citenamefont {Aprile}\ \emph {et~al.}(2018)\citenamefont {Aprile}
  \emph {et~al.}}]{Aprile:2018dbl}%
  \BibitemOpen
  \bibfield  {author} {\bibinfo {author} {\bibfnamefont {E.}~\bibnamefont
  {Aprile}} \emph {et~al.} (\bibinfo {collaboration} {XENON}),\ }\href
  {\doibase 10.1103/PhysRevLett.121.111302} {\bibfield  {journal} {\bibinfo
  {journal} {Phys. Rev. Lett.}\ }\textbf {\bibinfo {volume} {121}},\ \bibinfo
  {pages} {111302} (\bibinfo {year} {2018})},\ \Eprint
  {http://arxiv.org/abs/1805.12562} {arXiv:1805.12562 [astro-ph.CO]}
  \BibitemShut {NoStop}%
\bibitem [{\citenamefont {Cui}\ \emph {et~al.}(2017)\citenamefont {Cui} \emph
  {et~al.}}]{Cui:2017nnn}%
  \BibitemOpen
  \bibfield  {author} {\bibinfo {author} {\bibfnamefont {X.}~\bibnamefont
  {Cui}} \emph {et~al.} (\bibinfo {collaboration} {PandaX-II}),\ }\href
  {\doibase 10.1103/PhysRevLett.119.181302} {\bibfield  {journal} {\bibinfo
  {journal} {Phys. Rev. Lett.}\ }\textbf {\bibinfo {volume} {119}},\ \bibinfo
  {pages} {181302} (\bibinfo {year} {2017})},\ \Eprint
  {http://arxiv.org/abs/1708.06917} {arXiv:1708.06917 [astro-ph.CO]}
  \BibitemShut {NoStop}%
\bibitem [{\citenamefont {Akerib}\ \emph {et~al.}(2017)\citenamefont {Akerib}
  \emph {et~al.}}]{Akerib:2016vxi}%
  \BibitemOpen
  \bibfield  {author} {\bibinfo {author} {\bibfnamefont {D.~S.}\ \bibnamefont
  {Akerib}} \emph {et~al.} (\bibinfo {collaboration} {LUX}),\ }\href {\doibase
  10.1103/PhysRevLett.118.021303} {\bibfield  {journal} {\bibinfo  {journal}
  {Phys. Rev. Lett.}\ }\textbf {\bibinfo {volume} {118}},\ \bibinfo {pages}
  {021303} (\bibinfo {year} {2017})},\ \Eprint
  {http://arxiv.org/abs/1608.07648} {arXiv:1608.07648 [astro-ph.CO]}
  \BibitemShut {NoStop}%
\bibitem [{\citenamefont {Aprile}\ \emph
  {et~al.}(2019{\natexlab{a}})\citenamefont {Aprile} \emph
  {et~al.}}]{Aprile:2019bbb}%
  \BibitemOpen
  \bibfield  {author} {\bibinfo {author} {\bibfnamefont {E.}~\bibnamefont
  {Aprile}} \emph {et~al.} (\bibinfo {collaboration} {XENON}),\ }\href
  {\doibase 10.1103/PhysRevD.100.052014} {\bibfield  {journal} {\bibinfo
  {journal} {Phys. Rev.}\ }\textbf {\bibinfo {volume} {D100}},\ \bibinfo
  {pages} {052014} (\bibinfo {year} {2019}{\natexlab{a}})},\ \Eprint
  {http://arxiv.org/abs/1906.04717} {arXiv:1906.04717 [physics.ins-det]}
  \BibitemShut {NoStop}%
\bibitem [{\citenamefont {Aprile}\ \emph
  {et~al.}(2019{\natexlab{b}})\citenamefont {Aprile} \emph
  {et~al.}}]{Aprile:2019dme}%
  \BibitemOpen
  \bibfield  {author} {\bibinfo {author} {\bibfnamefont {E.}~\bibnamefont
  {Aprile}} \emph {et~al.} (\bibinfo {collaboration} {XENON}),\ }\href
  {\doibase 10.1103/PhysRevD.99.112009} {\bibfield  {journal} {\bibinfo
  {journal} {Phys. Rev.}\ }\textbf {\bibinfo {volume} {D99}},\ \bibinfo {pages}
  {112009} (\bibinfo {year} {2019}{\natexlab{b}})},\ \Eprint
  {http://arxiv.org/abs/1902.11297} {arXiv:1902.11297 [physics.ins-det]}
  \BibitemShut {NoStop}%
\bibitem [{\citenamefont {Aprile}\ \emph {et~al.}(2017)\citenamefont {Aprile}
  \emph {et~al.}}]{Aprile:2017aty}%
  \BibitemOpen
  \bibfield  {author} {\bibinfo {author} {\bibfnamefont {E.}~\bibnamefont
  {Aprile}} \emph {et~al.} (\bibinfo {collaboration} {XENON}),\ }\href
  {\doibase 10.1140/epjc/s10052-017-5326-3} {\bibfield  {journal} {\bibinfo
  {journal} {Eur. Phys. J.}\ }\textbf {\bibinfo {volume} {C77}},\ \bibinfo
  {pages} {881} (\bibinfo {year} {2017})},\ \Eprint
  {http://arxiv.org/abs/1708.07051} {arXiv:1708.07051 [astro-ph.IM]}
  \BibitemShut {NoStop}%
\bibitem [{\citenamefont {Mount}\ \emph {et~al.}(2017)\citenamefont {Mount}
  \emph {et~al.}}]{Mount:2017qzi}%
  \BibitemOpen
  \bibfield  {author} {\bibinfo {author} {\bibfnamefont {B.~J.}\ \bibnamefont
  {Mount}} \emph {et~al.},\ }\href@noop {} {\  (\bibinfo {year} {2017})},\
  \Eprint {http://arxiv.org/abs/1703.09144} {arXiv:1703.09144
  [physics.ins-det]} \BibitemShut {NoStop}%
\bibitem [{\citenamefont {Newstead}(2020)}]{zenodo}%
  \BibitemOpen
  \bibfield  {author} {\bibinfo {author} {\bibfnamefont {J.~L.}\ \bibnamefont
  {Newstead}},\ }\href {\doibase 10.5281/zenodo.3653516} {\enquote {\bibinfo
  {title} {Atmospheric neutrinos in a next-generation xenon dark matter
  experiment},}\ } (\bibinfo {year} {2020})\BibitemShut {NoStop}%
\bibitem [{\citenamefont {Chen}\ \emph {et~al.}(2017)\citenamefont {Chen},
  \citenamefont {Chi}, \citenamefont {Liu},\ and\ \citenamefont
  {Wu}}]{Chen:2016eab}%
  \BibitemOpen
  \bibfield  {author} {\bibinfo {author} {\bibfnamefont {J.-W.}\ \bibnamefont
  {Chen}}, \bibinfo {author} {\bibfnamefont {H.-C.}\ \bibnamefont {Chi}},
  \bibinfo {author} {\bibfnamefont {C.~P.}\ \bibnamefont {Liu}}, \ and\
  \bibinfo {author} {\bibfnamefont {C.-P.}\ \bibnamefont {Wu}},\ }\href
  {\doibase 10.1016/j.physletb.2017.10.029} {\bibfield  {journal} {\bibinfo
  {journal} {Phys. Lett.}\ }\textbf {\bibinfo {volume} {B774}},\ \bibinfo
  {pages} {656} (\bibinfo {year} {2017})},\ \Eprint
  {http://arxiv.org/abs/1610.04177} {arXiv:1610.04177 [hep-ex]} \BibitemShut
  {NoStop}%
\bibitem [{\citenamefont {Temples}\ \emph {et~al.}(2019)\citenamefont
  {Temples}, \citenamefont {Dahl}, \citenamefont {Lippincott}, \citenamefont
  {D.}, \citenamefont {Monte}, \citenamefont {McLaughlin},\ and\ \citenamefont
  {Phelan}}]{dylanTalk}%
  \BibitemOpen
  \bibfield  {author} {\bibinfo {author} {\bibfnamefont {D.}~\bibnamefont
  {Temples}}, \bibinfo {author} {\bibfnamefont {E.~C.}\ \bibnamefont {Dahl}},
  \bibinfo {author} {\bibfnamefont {H.~W.}\ \bibnamefont {Lippincott}},
  \bibinfo {author} {\bibfnamefont {B.}~\bibnamefont {D.}}, \bibinfo {author}
  {\bibfnamefont {A.}~\bibnamefont {Monte}}, \bibinfo {author} {\bibfnamefont
  {J.}~\bibnamefont {McLaughlin}}, \ and\ \bibinfo {author} {\bibfnamefont
  {J.}~\bibnamefont {Phelan}},\ }\href
  {http://www-kam2.icrr.u-tokyo.ac.jp/indico/event/3/session/10/contribution/414}
  {\enquote {\bibinfo {title} {Understanding neutrino background implications
  in lxe-tpc dark matter searches using 127xe electron captures},}\ } (\bibinfo
  {year} {2019}),\ \bibinfo {note} {tAUP}\BibitemShut {NoStop}%
\bibitem [{\citenamefont {Cowan}\ \emph {et~al.}(2011)\citenamefont {Cowan},
  \citenamefont {Cranmer}, \citenamefont {Gross},\ and\ \citenamefont
  {Vitells}}]{Cowan:2010js}%
  \BibitemOpen
  \bibfield  {author} {\bibinfo {author} {\bibfnamefont {G.}~\bibnamefont
  {Cowan}}, \bibinfo {author} {\bibfnamefont {K.}~\bibnamefont {Cranmer}},
  \bibinfo {author} {\bibfnamefont {E.}~\bibnamefont {Gross}}, \ and\ \bibinfo
  {author} {\bibfnamefont {O.}~\bibnamefont {Vitells}},\ }\href {\doibase
  10.1140/epjc/s10052-011-1554-0, 10.1140/epjc/s10052-013-2501-z} {\bibfield
  {journal} {\bibinfo  {journal} {Eur. Phys. J.}\ }\textbf {\bibinfo {volume}
  {C71}},\ \bibinfo {pages} {1554} (\bibinfo {year} {2011})},\ \bibinfo {note}
  {[Erratum: Eur. Phys. J.C73,2501(2013)]},\ \Eprint
  {http://arxiv.org/abs/1007.1727} {arXiv:1007.1727 [physics.data-an]}
  \BibitemShut {NoStop}%
\bibitem [{\citenamefont {Gonzalez-Garcia}\ \emph {et~al.}(2018)\citenamefont
  {Gonzalez-Garcia}, \citenamefont {Maltoni}, \citenamefont {Perez-Gonzalez},\
  and\ \citenamefont {Zukanovich~Funchal}}]{Gonzalez-Garcia:2018dep}%
  \BibitemOpen
  \bibfield  {author} {\bibinfo {author} {\bibfnamefont {M.~C.}\ \bibnamefont
  {Gonzalez-Garcia}}, \bibinfo {author} {\bibfnamefont {M.}~\bibnamefont
  {Maltoni}}, \bibinfo {author} {\bibfnamefont {Y.~F.}\ \bibnamefont
  {Perez-Gonzalez}}, \ and\ \bibinfo {author} {\bibfnamefont {R.}~\bibnamefont
  {Zukanovich~Funchal}},\ }\href {\doibase 10.1007/JHEP07(2018)019} {\bibfield
  {journal} {\bibinfo  {journal} {JHEP}\ }\textbf {\bibinfo {volume} {07}},\
  \bibinfo {pages} {019} (\bibinfo {year} {2018})},\ \Eprint
  {http://arxiv.org/abs/1803.03650} {arXiv:1803.03650 [hep-ph]} \BibitemShut
  {NoStop}%
\bibitem [{\citenamefont {Bœhm}\ \emph {et~al.}(2019)\citenamefont {Bœhm},
  \citenamefont {Cerdeño}, \citenamefont {Machado}, \citenamefont
  {Olivares-Del~Campo},\ and\ \citenamefont {Reid}}]{Boehm:2018sux}%
  \BibitemOpen
  \bibfield  {author} {\bibinfo {author} {\bibfnamefont {C.}~\bibnamefont
  {Bœhm}}, \bibinfo {author} {\bibfnamefont {D.~G.}\ \bibnamefont {Cerdeño}},
  \bibinfo {author} {\bibfnamefont {P.~A.~N.}\ \bibnamefont {Machado}},
  \bibinfo {author} {\bibfnamefont {A.}~\bibnamefont {Olivares-Del~Campo}}, \
  and\ \bibinfo {author} {\bibfnamefont {E.}~\bibnamefont {Reid}},\ }\href
  {\doibase 10.1088/1475-7516/2019/01/043} {\bibfield  {journal} {\bibinfo
  {journal} {JCAP}\ }\textbf {\bibinfo {volume} {1901}},\ \bibinfo {pages}
  {043} (\bibinfo {year} {2019})},\ \Eprint {http://arxiv.org/abs/1809.06385}
  {arXiv:1809.06385 [hep-ph]} \BibitemShut {NoStop}%
\bibitem [{\citenamefont {Dutta}\ \emph {et~al.}(2020)\citenamefont {Dutta},
  \citenamefont {Lang}, \citenamefont {Liao}, \citenamefont {Sinha},
  \citenamefont {Strigari},\ and\ \citenamefont {Thompson}}]{Dutta:2020che}%
  \BibitemOpen
  \bibfield  {author} {\bibinfo {author} {\bibfnamefont {B.}~\bibnamefont
  {Dutta}}, \bibinfo {author} {\bibfnamefont {R.~F.}\ \bibnamefont {Lang}},
  \bibinfo {author} {\bibfnamefont {S.}~\bibnamefont {Liao}}, \bibinfo {author}
  {\bibfnamefont {S.}~\bibnamefont {Sinha}}, \bibinfo {author} {\bibfnamefont
  {L.}~\bibnamefont {Strigari}}, \ and\ \bibinfo {author} {\bibfnamefont
  {A.}~\bibnamefont {Thompson}},\ }\href@noop {} {\  (\bibinfo {year}
  {2020})},\ \Eprint {http://arxiv.org/abs/2002.03066} {arXiv:2002.03066
  [hep-ph]} \BibitemShut {NoStop}%
\end{thebibliography}%

\end{document}